# STRUCTURAL AND MAGNETIC STUDIES ON La$_{2-x}$Dy$_x$Ca$_{2x}$Ba$_2$Cu$_{4+2x}$O$_z$ TYPE SUPERCONDUCTING OXIDES

S. Rayaprol[*] and Darshan C. Kundaliya
*Tata Institute of Fundamental Research, Mumbai – 400 005 (India)*
C. M. Thaker and D. G. Kuberkar
*Department of Physics, Saurashtra University, Rajkot – 360 005 (India)*
Keka R. Chakraborty, P. S. R. Krishna and M. Ramanadham
*SSPD, Bhabha Atomic Research Centre, Mumbai – 400 085 (India)*

**ABSTRACT**

The La-2125 type La$_{2-x}$Dy$_x$Ca$_{2x}$Ba$_2$Cu$_{4+2x}$O$_z$ (0.1 ≤ x ≤ 0.5; LDBO) compounds have been synthesized and studied for their structural and superconducting properties by room temperature neutron diffraction, high field dc magnetization, four-probe resistivity and iodometric double titration. The Rietveld analysis of the neutron diffraction data reveals tetragonal structure for all the samples, which crystallizes into La-123 type tetragonal structure in P4/mmm space group. Iodometric double titrations were performed to determine the oxygen content values and calculate mobile charge carrier (holes) density. The superconducting transition temperatures (T$_c$) increases from ~ 20 K for x = 0.1 to a maximum of 75 K for x = 0.5. Flux pinning force (F$_p$) and critical current density (J$_c$), calculated from the low temperature hysteresis loops, also increases with increasing dopant concentration. The paper presents the studies on structure and superconducting properties of all LDBO compounds in light of the role of calcium in inducing superconductivity in the tetragonal non-superconducting oxide.

1. **INTRODUCTION**

The mixed oxide superconductors having structure similar to the tetragonal REBa$_2$Cu$_3$O$_z$ (RE = Y or rare earth ion), has attracted interest since several RE-123 type tetragonal structures such as REBaCaCu$_3$O$_{7-\delta}$ [1], LaBa$_{1.5}$Ca$_{0.5}$Cu$_3$O$_\delta$ [2], La$_{2-x}$RE$_x$Ca$_{2x}$Ba$_2$Cu$_{4+2x}$O$_z$ [3], (Ca$_x$La$_{1-x}$)(Ba$_{c-x}$La$_{2-c+x}$)Cu$_3$O$_y$ [4], La$_{3.5-x-y}$RE$_y$Ca$_{2x}$Ba$_{3.5-x}$Cu$_7$O$_z$ [5] etc, have been studied extensively for their structural and superconducting properties. These tetragonal superconductors are especially interesting because

(i) The superconducting transition temperature (T$_c$) largely depends on the cationic intermixing at various atomic sites unlike in the orthorhombic RE-123 superconductors where the T$_c$ depends on the optimal value of oxygen content (z ~ 6.9). For the superconducting tetragonal compounds, the oxygen content value per unit cell (z) can be > 7, which is in the 'over hole doped' region.

(ii) The crystal structure of tetragonal superconductors remains tetragonal throughout the doping range, thus avoiding the complications in interpreting experimental data, arising from the ordered Cu-O chains which is observed for the orthorhombic RE-123 superconductors, on structural (orthorhombic-tetragonal) transition.

Some of us have studied the La-2125 ($La_2Ca_1Ba_2Cu_5O_z$) compounds in the stoichiometric form $La_{2-x}RE_xCa_{2x}Ba_2Cu_{4+2x}O_z$ with x varying from 0 to 0.5. The La-2125 phase is obtained for x = 0.5, and it exhibits maximum $T_c \sim$ 78 ($\pm$1) K. The $La_{2-x}RE_xCa_{2x}Ba_2Cu_{4+2x}O_z$ with x = 0.1 – 0.5 type high temperature superconductors has been studied for different $R^{3+}$ ions such as Nd, Gd, Pr etc [6-8].

The starting compound of this family La-224 ($La_2Ba_2Cu_4O_z$; for x = 0.0 in $La_{2-x}RE_xCa_{2x}Ba_2Cu_{4+2x}O_z$), is a tetragonal triple perovskite compound exhibiting semi-conducting behavior [6]. La-224 can be normalized to $RBa_2Cu_3O_{7-\delta}$ (RE-123) form as $La_1(Ba_{1.5}La_{0.5})Cu_3O_{7-\delta}$. The structural analysis of the neutron diffraction data on La-224 exhibits that La-224 is isostructural to the tetragonal RE-123 structure in P4/mmm space group [6, 7]. The bulk superconductivity in this tetragonal compound is achieved when $Ca^{2+}$ is doped at $La^{3+}$ site in the nominal composition. In the entire doping range structure remains similar to that of tetragonal $RBa_2Cu_3O_{6+\delta}$ (P4/mmm) [9-10]. The $T_c$ increases up to 75 K ($\pm$1) for an optimum value of x = 0.5. The structure of a typical La-2125 unit cell is shown in the Figure 1. The figure shows the atomic site occupancies of various cations. $La^{3+}$ site is occupied by $La^{3+}$, $Ca^{2+}$ and substituted $R^{3+}$, where as the $Ba^{2+}$ site is occupied by $La^{3+}$ and remaining $Ca^{2+}$. This type of cationic intermixing is responsible for the superconductivity in La-2125 type compounds. We have observed that the 'hole doping' by $Ca^{2+}$ at $La^{3+}$ is not fully compensated by 'hole filling' by $La^{3+}$ at $Ba^{2+}$ site, hence the increase in hole concentration, and $T_c$.

The mobile charge carriers (i.e., *holes*), which are responsible for superconductivity in this compound, are introduced by the substitution of $Ca^{2+}$ for $La^{3+}$. These mobile p-type carriers first diminish the long range magnetic ordering of Cu-spins in the tetragonal non-superconductors like La-224, and drive the system from antiferromagnetic insulator to a metallic superconductor [11-13]. It is known that $Ca^{2+}$ substitution creates oxygen vacancies in $Cu-O_2$ planes in $(Y/R)_{1-x}Ca_xBa_2Cu_3O_z$ (1-2-3 type system) [14]. The possibility of oxygen vacancies in $Cu-O_2$ planes is of interest considering that superconductivity is essentially believed to arise in $Cu-O_2$ planes in high $T_c$ materials. The high temperature superconductivity in the tetragonal 1-2-3 and similar type of phases suggest that, the Ca cations introduce holes into the bridging site between two $Cu-O_2$ sheets. This alternative "bridging" hole picture, as suggested by H. Gu et al [15] for $LaBaCaCu_3O_{7-\delta}$ compounds, is more general in the tetragonal high $T_c$ cuprates than the "transferring" picture associated with the orthorhombic $RBa_2Cu_3O_{7-\delta}$ phases.

Among the RE-123 derived high $T_c$ phases, we present here our studies on $La_{2-x}Dy_xCa_{2x}Ba_2Cu_{4+2x}O_z$ (LDBO) with structural and superconducting property considerations. In this paper we discuss the role of calcium in introducing superconductivity in LDBO compounds by 'bridging' the $CuO_2$ sheets.

2   **EXPERIMENTAL**

All samples in the LDBO series were prepared by solid-state reaction method. The high purity (99.9+ % pure) starting compounds of $La_2O_3$, $Dy_2O_3$, $CaCO_3$, $BaCO_3$ and CuO were taken in stoichiometric quantities and ground thoroughly under acetone to form a homogeneous mixture. This mixture was then calcined at $900^0C$ for 24 hours in powder form. The samples were then palletized and sintered at $940^0$C for 48 hours with intermediate grindings. These sintered pellets were first annealed in flowing $N_2$ gas at $940^0$C for 12 hours followed by $O_2$ gas annealing for another 12 hours. While cooling, the samples were first slow cooled to $500^0$C at $1^0$C/min rate and annealed for 24 hours at this temperature and then finally slow cooled to room temperature in flowing $O_2$ at $1^0$C/min rate. The samples were characterized for phase purity and lattice constants by room temperature X-ray diffraction. All the samples were found to be single phase with tetragonal symmetry. The four-probe resistivity setup was used to determine the superconducting transition temperature ($T_c$) using a closed cycle cryocooler. The lowest temperature, which we could achieve with this system, was 40 K, and hence the resistance was measured up to that temperature. Quantum design SQUID magnetometer was used to measure dc susceptibility ($\chi$) and dc magnetization values. Oxygen content was determined by Iodometric double titration method. Powdered samples of all the compositions (~ 7 gm) were used for Neutron diffraction. The samples were packed in Vanadium cans and exposed to the monochromatic neutron beam ($\lambda$ = 1.249 Å). The neutron diffraction data was recorded on a 'Position sensitive detector'. The analysis of the neutron data has been done using FULLPROF suite employing Rietveld refinement method [16-17].

## 3.1 RESULTS

### 3.1.1 Structure

The neutron diffraction experiments were performed on the polycrystalline powders of LDBO compounds. The structural analysis reveals tetragonal La-123 type structure with P4/mmm space group for all samples studied in the x = 0.1 – 0.5 range. The detailed analysis of neutron diffraction data on $La_{2-x}Dy_xCa_{2x}Ba_2Cu_{4+2x}O_z$ (with x = 0.1 – 0.5) by Rietveld refinement method has been reported in Ref 18. A typical unit cell of the La-2125 compound is shown in the Figure 1.

Using the BONDSTR module of the FULLPROF suite, we have calculated the bond lengths for all the compositions studied. The variation of selected bond lengths, around the conducting $CuO_2$ plane, is plotted as a function of dopant concentration in Figure 2. The increase in content of smaller ionic radii ions ($Ca^{2+}$, $Dy^{3+}$ at $La^{3+}$ site and $Ca^{2+}$, $La^{3+}$ at $Ba^{2+}$) results in the shrinking of these bond lengths thus distorting the polyhedra of the perovskite block containing the superconducting $CuO_2$ planes. Due to the shortening of the bond lengths the unit cell parameters also decrease, in similar fashion, resulting in the reduction of the unit cell volume, resulting in the bridging of the two $CuO_2$ planes.

### 3.1.2 Resistance and susceptibility measurements

The superconducting properties of all the LDBO samples were determined by electrical (four probe resistance measurement, R) and magnetic (dc susceptibility, $\chi$) methods.

Figure 3 shows plot of R and $\chi$ as a function of temperature for all the samples of LDBO system in the temperature range 5 – 100 K. The $T_c$ values obtained by these independent methods are in good agreement. Since the $T_c^{R=0}$ for x = 0.1 and 0.2 samples were below 40 K, we determined the transition temperatures for all samples precisely from the first derivative of the susceptibility curves. With the increasing concentration *x*, the superconducting transition temperature increases from ~ 20 K for x = 0.1 to maximum of $T_c^{R=0}$ = 75 K for x = 0.5 (La-2125 phase) compound.

The values of transition temperatures obtained from both resistivity and susceptibility measurements are given in Table 1. $T_c^{ON}$ has been plotted as a function of increasing 'x' in Figure 4.

### 3.1.3 *Oxygen content determination and charge carrier density*

Iodometric double titration has been performed on all the samples to calculate the oxygen content per unit formula. Using the oxygen content values obtained from Iodometric titrations and Tokura's formula [19-20], hole concentrations have been calculated. The values of hole concentration, oxygen content per unit cell and per unit formula are tabulated in Table 1.

The increase in the transition temperature with *x* can also be compared with the simultaneous increase of hole concentration in sheets ($p_{sh}$) with *x*, as clearly seen in Figure 4. The simultaneous increase in $p_{sh}$ and $T_c$ with increasing *x* indicates towards the correlation between the hole concentration and superconducting properties for these compounds.

### 3.1.3 *Magnetization and determination of critical current density*

Higher critical current density in the superconductors can be achieved by blocking the movement of the flux lines. This can be achieved by the '*defects*'. The bridging of the $CuO_2$ sheets by $Ca^{2+}$ is accompanied by the creation of '*flux pinning centers*', as structural defects are created when smaller ion replaces a bigger ion in the unit cell. We have therefore measured critical current density ($J_c$) for all the samples at 5K, and observed increase in $J_c$ with increasing dopant concentration. Figure 5 shows moment (in emu) plotted again the sweeping field. These M-H loops were used to calculate critical current density ($J_c$ expressed in A/cm$^2$) given by Bean's critical state model and expressed as

$$J_c = \frac{30[M^+ - M^-]}{D} \quad \text{---(1)}$$

where $M^+$ and $M^-$ are the values of magnetization observed during the up and down cycle of the field sweeping [21-22]. *D* is the average grain size calculated from the SEM micrographs has been taken ~ 2 x 10$^{-4}$ cm. Figure 6 shows the variation of $J_c$ with applied field. With increasing x, the $J_c$ increases up to x = 0.4 and then decreases slightly. Thus x = 0.4 can be taken as the optimum doping level up to which the $J_c$ increases and enhances flux pinning ($F_p$) which is the product of $J_c$ and H at a particular temperature.

The $Dy^{3+}$ moment in the low temperature has no influence on the superconducting properties. This is obvious because in the La-2125 unit cell La-site is far removed from the Cu-O plane and therefore the magnetic moment has no influence.

## *3.2* DISCUSSION

Superconductivity in the orthorhombic RE-123 oxides is manifested by the 'transfer of holes' from *charge reservoir* Cu-O chain to the conductive Cu-O$_2$ sheets, but the conduction of charge carriers between two conducting CuO$_2$ layers in tetragonal Ca doped RE-123 phases should be different since there are no chains in the tetragonal 1-2-3 phase. The Ca-cations at RE site in RE-123 systems introduce necessary holes between the two CuO$_2$ sheets which couple the conductive sheets in the tetragonal 1-2-3 phases which is called the 'bridging' of the CuO$_2$ sheets.

The substitution of Ca$^{2+}$ at La$^{3+}$ leads to the creation of hole at La$^{3+}$ site which bridges the two CuO$_2$ sheets resulting in the '*turning on*' superconductivity in the presently studied LDBO (La-2125 type) system. The results of Rietveld analysis show the increase in Ca$^{2+}$ content at La$^{3+}$ site, which is reflected in an increase in the hole concentration. This, in LDBO system, is supported by the iodometric titration results also (Figure 4). Thus, we can confidently attribute the increase in superconducting transition temperature to the increase in the concentration of Ca$^{2+}$ at La$^{3+}$ which helps in increasing the mobile charge carrier concentration.

The local charge environment in the CuO – LaO – CuO layer, (the perovskite block) gets affected by Ca doping. The bridging of the CuO$_2$ planes by Ca$^{2+}$ substitution leads to the shortening of the bond lengths in the perovskite block and as a result of which the unit cell volume also decreases. Defects introduced in the Cu-O$_2$ planes due to substitution of Ca$^{2+}$ for La$^{3+}$ (i.e., an extra hole) results in the conduction of charge carriers through this bridged path. The bridging of two Cu-O layers causes buckling of the conducting plane and results in the creation of centers for flux trapping, which can be attributed due to the simultaneous increase in T$_c$, J$_c$ and F$_p$ with increasing dopant concentration.

## *4* CONCLUSION

Superconductivity is induced in a non-superconducting La$_2$Ba$_2$Cu$_4$O$_z$ phase by simultaneous addition of CaO and CuO alongwith Dy$^{3+}$ substitution. The role of Dy$^{3+}$ has been found to stabilize the crystal structure, and has no pronounced effect on the superconducting transition temperature. Ca$^{2+}$ substitution helps in inducing superconductivity by creating holes at La$^{3+}$ site resulting in the bridging of the two conductive CuO$_2$ sheets. The maximum T$_c$, which has been achieved for LDBO system, is 75K, and can be correlated with the optimum hole concentration value. We observe here that Ca$^{2+}$ plays a significant role in introducing superconductivity in La$_{2-x}$Dy$_x$Ca$_{2x}$Ba$_2$Cu$_{4+2x}$O$_z$ system and creating 'pinning centers' by the addition of extra holes, which results in the increase of critical current density and flux pinning but without inducing any structural transition

**Table 1** Values of $T_c$ obtained from resistivity and susceptibility measurement, oxygen content from Iodometric titration and hole concentration per unit cell and in Cu-$O_2$ sheets

| Dopant concentration | Transition temperature | | Oxygen content (in 123 & 2125) | | Hole concentration | |
|---|---|---|---|---|---|---|
| x | $T_c^{R=0}$ (K) (± 1 K) | $T_c^{\chi}$ (Onset) (± 1 K) | per unit cell (z') | per unit formula (z) | per unit cell (p) | in $CuO_2$ sheets ($p_{sh}$) |
| 0.1 | < 20 | 38 | 6.90 (2) | 9.66 (2) | 0.266 (2) | 0.200 (2) |
| 0.2 | < 30 | 58 | 6.95 (2) | 10.19 (2) | 0.300 (2) | 0.225 (2) |
| 0.3 | 66 | 68 | 6.96 (2) | 10.68 (2) | 0.306 (2) | 0.230 (2) |
| 0.4 | 71 | 78 | 6.96 (2) | 11.14 (2) | 0.309 (2) | 0.232 (2) |
| 0.5 | 75 | 79 | 6.97 (2) | 11.61 (2) | 0.313 (2) | 0.235 (2) |

Figure 1    A typical La-2125 unit cell.

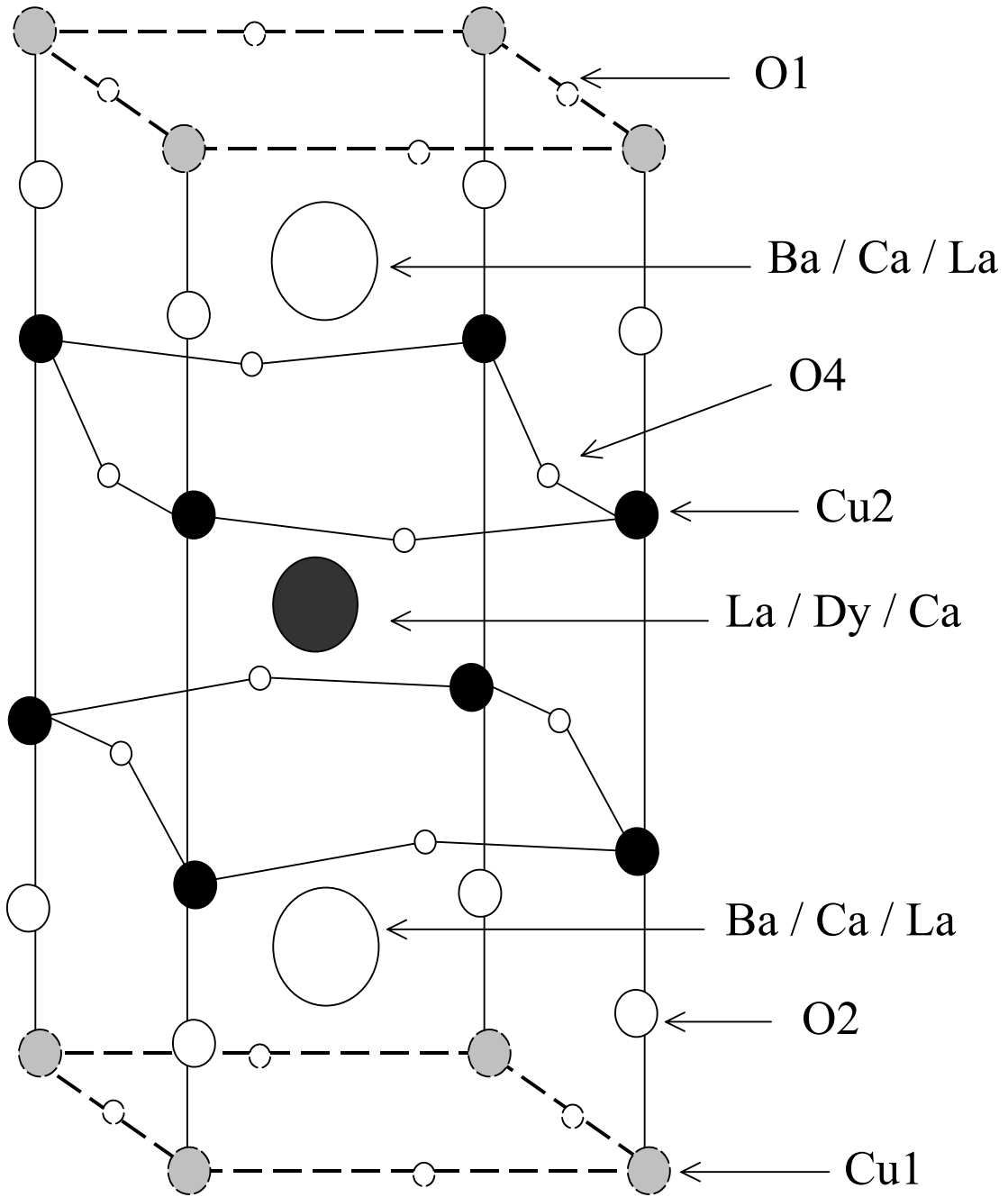

Figure 2   Variation of (selected) bond lengths and lattice parameters with increasing dopant concentration for $La_{2-x}Dy_xCa_{2x}Ba_2Cu_{4+2x}O_z$ series.

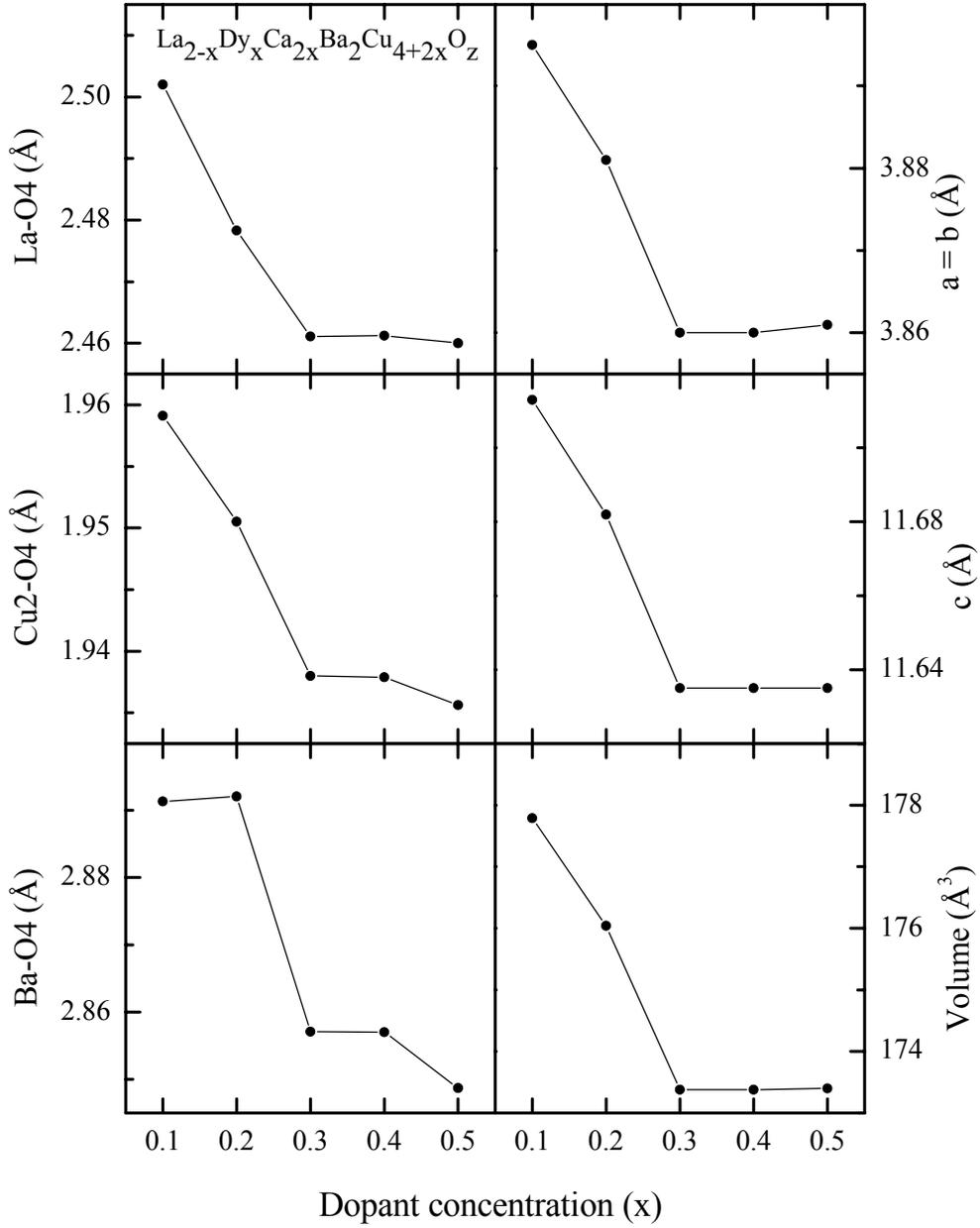

Figure 3   The combined plot of resistance and susceptibility as a function of temperature for all samples in $La_{2-x}Dy_xCa_{2x}Ba_2Cu_{4+2x}O_z$ series.

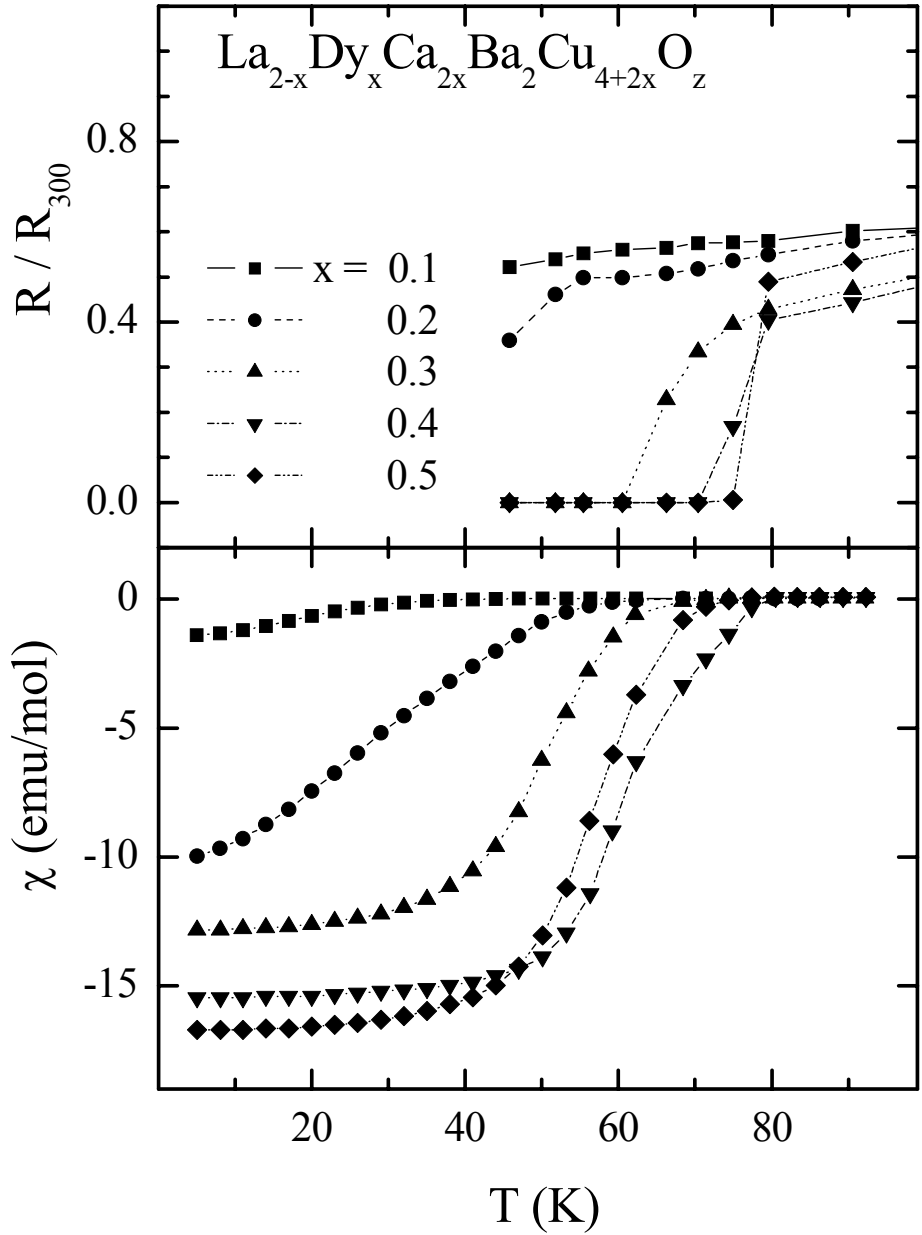

Figure 4  Increase in superconducting transition temperature ($T_c^{ON}$) and hole concentration (p) with increasing dopant concentration in $La_{2-x}Dy_xCa_{2x}Ba_2Cu_{4+2x}O_z$

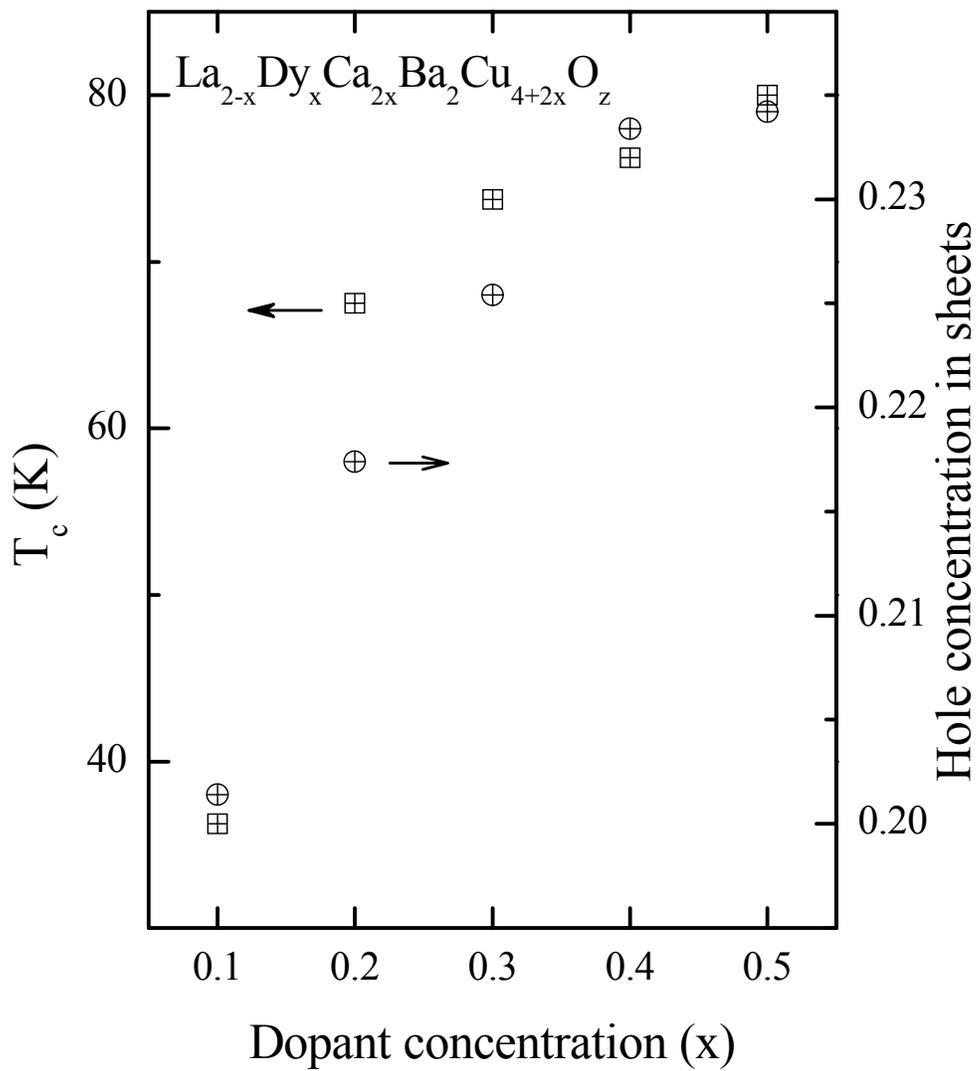

Figure 5    Hysteresis loops for $La_{2-x}Dy_xCa_{2x}Ba_2Cu_{4+2x}O_z$ (x = 0.1 – 0.5) samples measured at 5 K

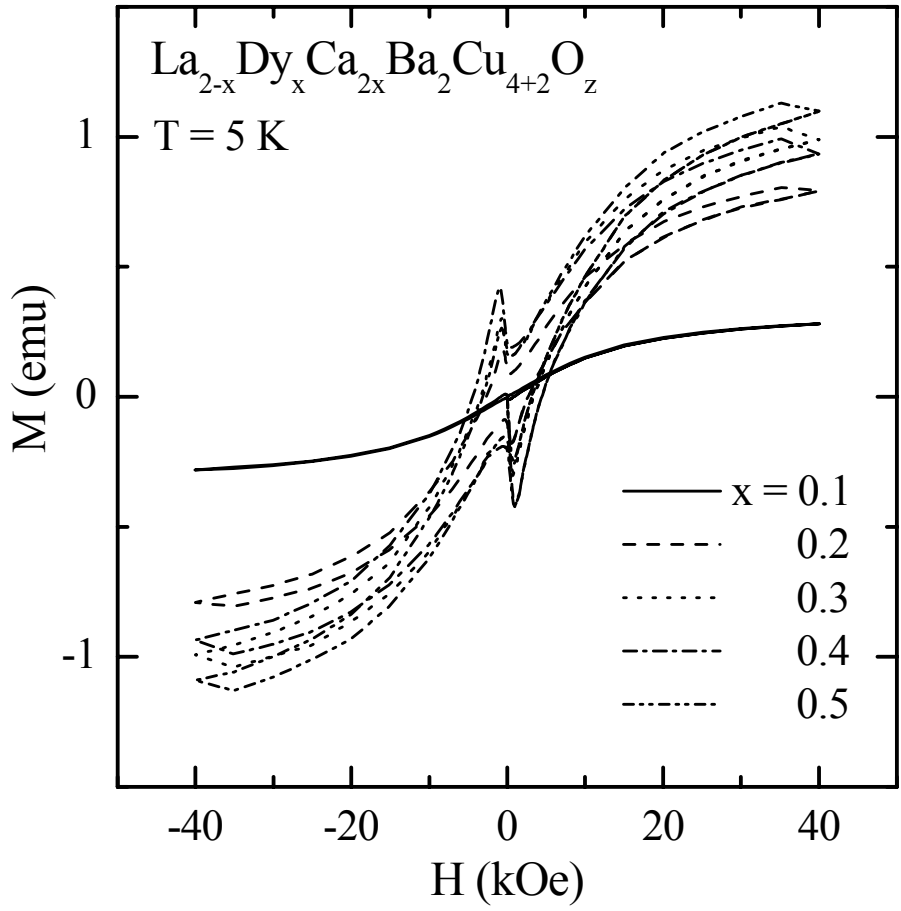

Figure 6    Critical current density ($J_c$) as a function of varying field for
$La_{2-x}Dy_xCa_{2x}Ba_2Cu_{4+2x}O_z$ (x = 0.1 – 0.5) compounds calculated at 5 K.
The insert shows $J_c$ vs. H for x = 0.1 sample.

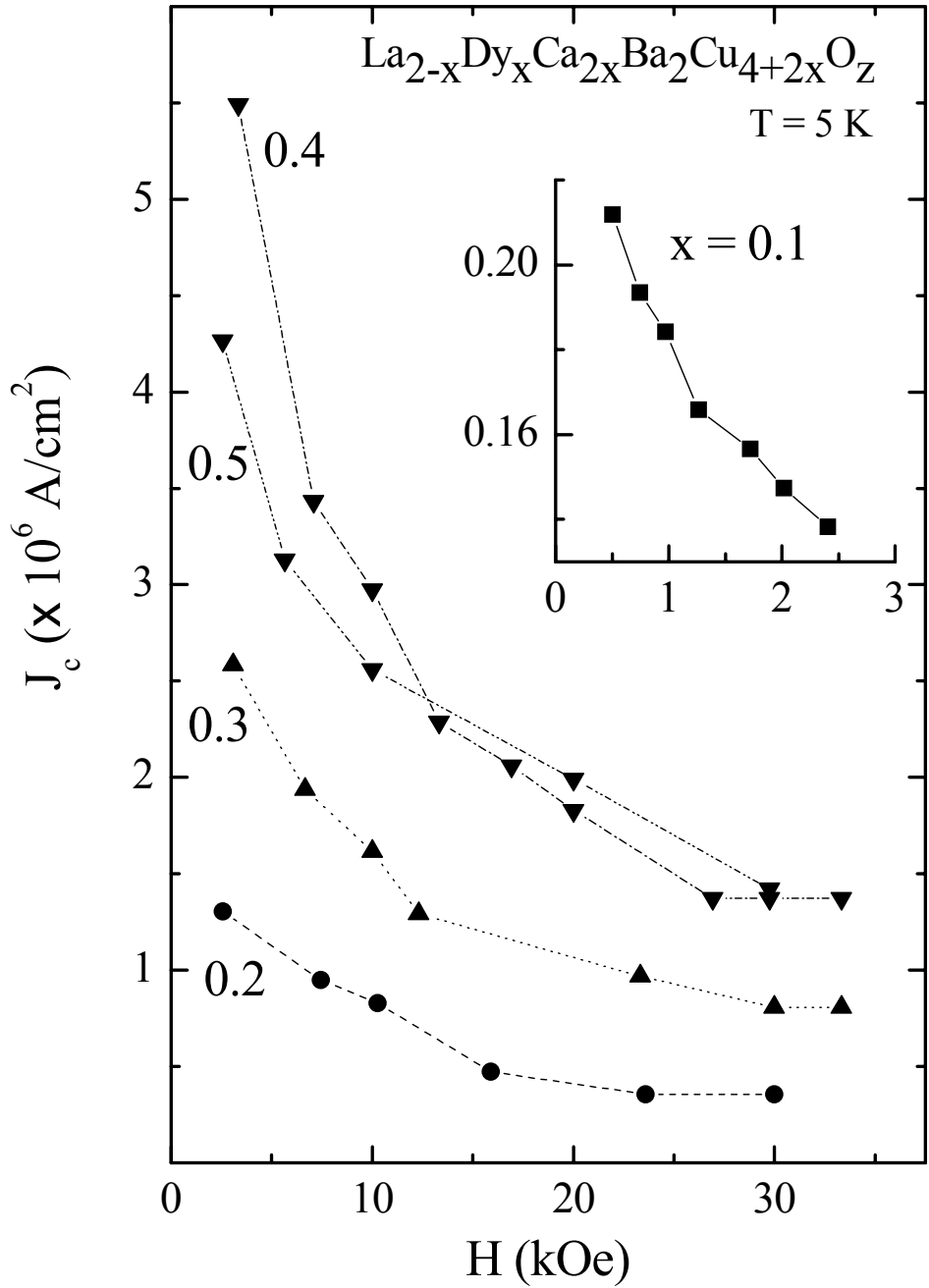